\documentclass[twocolumn,floatfix,pra,aps,showpacs]{revtex4}
\usepackage{epsfig,graphicx,tabularx}
\usepackage{amsmath}

\begin{document}

\title{Energy barriers for vortex nucleation in dipolar condensates}

\author{M. Abad}
\email{vinyet@ecm.ub.es}
\affiliation{Departament d'Estructura i Constituents de la Mat\`{e}ria,\\
Facultat de F\'{\i}sica, and IN2UB, Universitat de Barcelona, E--08028 Barcelona, Spain}
\author{M. Guilleumas}
\affiliation{Departament d'Estructura i Constituents de la Mat\`{e}ria,\\
Facultat de F\'{\i}sica, and IN2UB, Universitat de Barcelona, E--08028 Barcelona, Spain}
\author{R. Mayol}
\affiliation{Departament d'Estructura i Constituents de la Mat\`{e}ria,\\
Facultat de F\'{\i}sica, and IN2UB, Universitat de Barcelona, E--08028 Barcelona, Spain}
\author{M. Pi}
\affiliation{Departament d'Estructura i Constituents de la Mat\`{e}ria,\\
Facultat de F\'{\i}sica, and IN2UB, Universitat de Barcelona, E--08028 Barcelona, Spain}
\author{ D. M. Jezek   }
\affiliation{Departamento de F\'{\i}sica, Facultad de Ciencias Exactas
y Naturales,
Universidad de Buenos Aires, RA-1428 Buenos Aires, Argentina}
\affiliation{Consejo Nacional de Investigaciones Cient\'{\i}ficas y
T\'ecnicas, Argentina}

\date{\today}
\begin{abstract}
We consider singly-quantized vortex states in a condensate of $^{52}$Cr atoms in a pancake trap. We obtain the vortex solutions by
numerically solving the Gross-Pitaevskii equation in the rotating frame with
no further approximations.  The behavior of the condensate is studied under three different situations concerning the interactions: only $s$-wave, $s$-wave plus dipolar and only dipolar. The energy barrier for
the nucleation of a vortex is calculated as a
function of the vortex displacement from the rotation axis in the three cases. These results are compared to those obtained for contact interaction condensates in the Thomas-Fermi approximation, and to a pseudo-analytical model, showing this latter a very good agreement with the numerical calculation. 

\end{abstract}
\maketitle
\section{Introduction}\label{Intro}

%
%
%

Chromium condensates were first experimentally realized in 2005 \cite{gri05}. In contrast to condensates of alkali atoms, $^{52}$Cr condensates present an additional, non-negligible interaction, the dipolar interaction. It is non-local, anisotropic and long-range and these features introduce novel aspects in the physics of Bose-Einstein condensates (BECs) \cite{Santos2000,Lahaye2007}. 
The atom-atom interaction
is then determined by the balance of the $s$-wave contact interaction and the dipolar interaction, allowing thus to investigate the effects of their interplay. Although alkali atoms also possess some magnetic dipole moment, it is thirty-six times weaker than in chromium, which makes the latter a suitable system to experimentally study dipolar condensates of atomic gases \cite{gri05, Beaufils2008}. 

The evidence of the superfluid character of BECs comes mainly from the appearance of singly-quantized vortices above a critical rotation frequency as a means to convey angular momentum to the system. Vortices have not been experimentally realized in dipolar BECs yet, but they are a most appealing phenomenon and there is intensive theoretical research on them \cite{Yi2006,dell07,Wilson2009,Abad2009,Bijnen2009,Klawunn,LLL}.

The critical frequency for vortex nucleation in dipolar condensates has been predicted theoretically in different papers \cite{dell07,Abad2009}.
This critical frequency is based on a thermodynamical consideration that takes into account the energy of the vortex at the center of the trap \cite{Dalfovo1999}. However, in the present understanding of vortex formation (see Ref.~\cite{Fetter2009} for a review), the nucleation process takes place at the surface of the condensate and then the vortex enters it. Therefore, the theoretical prediction provides a lower bound to the experimental critical frequency \cite{Madison2000, AboShaeer2001}.
This difference can be understood in terms of an energy barrier the system needs to overcome to bring the vortex from the surface to the center of the condensate \cite{Svidzinsky2000,Kramer2002}.
That is, the nucleation of a vortex is associated with the existence of an
energy barrier in the configuration space between the initial
vortex-free state and the final centered vortex state. 
 The formation energy of a vortex can be estimated by calculating the
energy of a single off-center vortex 
as a function of the vortex core position.

The aim of this work is to quantify this extra energy that should be given to dipolar condensates to nucleate a vortex. Since we are interested mainly on the effects of the dipolar interaction, we will consider a very small $s$-wave scattering length. In the numerical calculation, we solve the three-dimensional (3D) Gross-Pitaevskii equation in the rotating frame using the imaginary-time evolution method. We obtain the vortex states, the vortex formation energy and hence the nucleation barrier.

This work is organized as follows. In Sec.~\ref{Theory} we describe the
theoretical framework and the system under study.
In Sec.~\ref{TF} the analytic expressions for the energy barrier in the Thomas-Fermi regime are given in 2D and 3D.
In Sec.~\ref{Numerical} we present the numerical calculation of the nucleation barrier and compare it both to the Thomas-Fermi results for purely $s$-wave condensates and to a pseudo-analytical model which is shown to reproduce very well the numerical results.
Finally, a summary and concluding remarks are
offered in Sec.~\ref{Conclusions}.

\section{Theoretical framework}\label{Theory}

Weakly interacting dipolar condensates are well described in the mean-field regime by the Gross-Pitaevskii (GP) equation, which includes a new term taking into account the dipole-dipole interaction.
%
%
In this framework it is possible to numerically study vortex states by working in the rotating frame \cite{Fetter2009}, that is a reference frame that rotates at a frequency $\Omega$ around the $z$ axis. The GP equation then reads
\begin{eqnarray}
 &&
 \left[ -\frac{ \hbar^2}{2m} \nabla^2 + V_{\text{trap}}(\mathbf{r}) +
g \, |\psi(\mathbf{r})|^2 \right. + \left. V_{\text{dip}}(\mathbf{r})
-\Omega \hat{L}_z\right] \psi(\mathbf{r}) \nonumber \\
&& 
 \qquad\qquad= \tilde{\mu} \, \psi(\mathbf{r}) \,,
\label{gp}
\end{eqnarray}
where $m$ is the atomic mass, $\psi(\mathbf{r})$ is the
condensate wave function normalized to the total number of
particles, $\hat{L}_z$ is the angular momentum operator along the $z$ axis and $\tilde{\mu}$ is the chemical potential. The terms in the left-hand side of the equation are, respectively, the kinetic energy, the trapping potential, the $s$-wave contact interaction, the dipolar potential and the rotating potential. When the dipolar BEC is at rest, the vortex-free state is obtained from
Eq.~(\ref{gp}) by setting $\Omega=0$.

The trapping potential is assumed to be harmonic and with axial symmetry, being $z$ the symmetry axis
\begin{equation}
V_{\text{trap}}(\mathbf{r})=
  \frac{m}{2}\, (\omega_{\perp}^2 r_\perp^2 +\omega_{z}^2 z^2) \,,
\end{equation}
where $r_\perp^2=x^2+y^2$ and $\omega_{\perp}$
and $\omega_{z}$ are the radial and axial angular trap frequencies,
respectively.

The contact interaction potential is characterized by the
coupling constant $g=4\pi\hbar^2 a /m$, with $a$ the $s$-wave scattering length. Using the physics of Feshbach resonances, the value of $a$ can be tuned to very small values by changing the magnitude of an external magnetic field \cite{Feshbach}. If this tuning is performed far enough from the actual resonance, the condensate is not destroyed and experiments can be indeed carried out \cite{Koch2008}.

The mean-field dipolar interaction $V_{\text{dip}} (\mathbf{r})$ is given by the integral
\begin{equation}
 V_{\text{dip}} (\mathbf{r})= \int d\mathbf{r'} v_{\text{dip}} (\mathbf{r}-\mathbf{r'}) |\psi(\mathbf{r'})|^2\ , \label{Vdip}
\end{equation}
where
\begin{equation}
v_{\text{dip}} (\mathbf{r}-\mathbf{r'})= \frac{\mu_0 \mu^2}{4 \pi}
\frac{1 - 3 \cos^2 \theta}{|\mathbf{r}-\mathbf{r'}|^3} 
 \label{dip-pot}
\end{equation}
is the dipole-dipole potential. In this equation $\mu$ is the magnetic dipole moment of the atoms, $\mu_0$ is the vacuum permeability,
$\mathbf{r}-\mathbf{r'}$ is the distance between the dipoles, and
$\theta$ is the angle between
the vector $\mathbf{r}-\mathbf{r'}$ and the dipole axis, which we also take to be $z$. Note that dipolar interactions are anisotropic, being attractive when the dipoles are head-to-tail and repulsive when they are side-by-side. Therefore, in our symmetry configuration, the dipolar interaction will be attractive along $z$ and repulsive in $x$ and $y$ directions. Since we are dealing with pancake-like condensates, the overall interaction will be mainly repulsive, although the small attractive part is what can bring instability and collapse to the condensate \cite{Santos2003,Ronen2007,Dutta2007,Bijnen2007,Lahaye2008}.

When the rotation frequency $\Omega$ exceeds a critical frequency $\Omega_c$, the centered-vortex state becomes energetically favorable and Eq.~(\ref{gp}) converges to such a solution, irrespective of the initial conditions. However, an off-axis vortex is not a stable state of the system and therefore cannot be found as a solution of GP equation as it appears in (\ref{gp}). One thus needs to impose some constraint during the minimization process to fix the vorticity along a line $(x_v, y_v,z)$. In this work this is achieved as follows. First, we generalize the Feynmann-Onsager ansatz for a centered-vortex state to the case of an off-axis vortex \cite{Abad2009},
\begin{equation} \psi(\mathbf{r})= \psi_0(\mathbf{r})\frac{x-x_v+i(y-y_v)}{\sqrt{(x-x_v)^2 + (y-y_v)^2}} \  ,\label{EqVortex2}
\end{equation}
and use it as the initial wave function. Here $\psi_0(\mathbf{r})$ is the ground state vortex-free wave function obtained directly from Eq.~(\ref{gp}) with $\Omega=0$.
This expression fixes a unit of vorticity along the line $(x_v,y_v,z)$ and corresponds to a straight vortex line. To maintain such a vorticity during the imaginary-time minimization we impose the conditions \cite{Abad2009}
\begin{align}
 \text{R}\text{e}[\psi(x_v,y,z)]&= 0\quad\forall y,z \\
 \text{I}\text{m}[\psi(x,y_v,z)]&= 0\quad\forall x,z 
\end{align}
to the wave function.
These relations are the equivalent of considering the vortex line as the intersection of two flat nodal surfaces, instead of allowing them to have some curvature. 
With this method, the quantization of the circulation is ensured in all cases, but the solutions are restricted to the case of straight vortex lines. 

In order to compute the energy of the condensate we evaluate the energy density functional in the rotating frame, which has the standard GP form but
with a new term, $E_{\text{dip}}$, which is the interaction energy
due to the dipole-dipole potential,
\begin{eqnarray}
 E [\psi]  && = E_{\text{kin}} + E_{\text{trap}} + E_{\text{int}} + E_{\text{dip}} + E_L= \nonumber\\
 && \hspace{-0.8cm}= \int\frac{ \hbar^2 }{2 m}  |\nabla \psi |^2 d {\bf r} + \int V_{\text{trap}}(\mathbf{r}) \,|\psi|^2 d {\bf r} + \int\frac{g}{2} \, |\psi|^4 d {\bf r} +   \nonumber\\
&& \hspace{-0.8cm} + \frac{1}{2} \int V_{\text{dip}}(\mathbf{r}) \,
|\psi|^2 d\mathbf{r}\, - \Omega\int\psi^* \hat{L}_z\psi\, d\mathbf{r}.
\label{ed}
\end{eqnarray}

In this work we will consider a $^{52}$Cr BEC of $N=10^5$ atoms in a trap with frequencies $\omega_\perp=8.4\times2\pi$ s$^{-1}$ and $\omega_z=98.5\times2\pi$ s$^{-1}$, with a magnetic dipole moment of $\mu=6\,\mu_B$ and a scattering length $a=5\,a_B$. The oscillator length in the plane, $a_\perp=\sqrt{\hbar/m\omega_\perp}$, gives an order of magnitude of the size of the condensate, and in our case takes the value $a_\perp=4.8\,\mu\text{m}$. By evolving the GP equation (\ref{gp}) in imaginary-time until convergence is reached, we find its solution. This is equivalent to a minimization of the energy density functional (\ref{ed}). 

\begin{figure}
 \epsfig{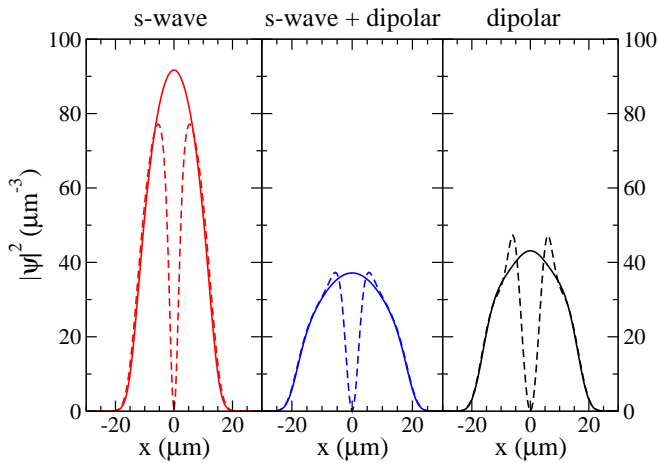}
 \caption{Density profiles for vortex (dashed) and ground (solid) states of condensates whose atoms interact via $s$-wave (left panel), $s$-wave plus dipolar (middle panel) and dipolar (right panel) interactions.}\label{profiles}
\end{figure}

In Fig.~\ref{profiles} we show the density profiles of the ground state and the singly-quantized vortex state of such a condensate in the three situations: when only $s$-wave interactions are considered, when $s$-wave plus dipolar interactions are considered, and when only dipolar interactions are considered. We see there that the dipolar interactions make the condensate more repulsive, hence increasing its size in $x$ and $y$ directions. For more detailed discussions of condensate deformation due to anisotropic dipolar interactions see Refs.\cite{Abad2009,Stuhler2005,Griesmaier2007} and references therein.

\section{Thomas-Fermi models for the energy barrier}\label{TF}

In purely $s$-wave condensates, when the number of particles is large and the repulsive contact interaction is strong, surface effects at the boundary of the condensate become negligible. Under these conditions one can neglect the kinetic energy term in Eq.~(\ref{gp}) and show that the density has a parabolic profile. This allows to find analytical solutions for the BEC. This approximation is known as Thomas-Fermi approximation (TF) and can be applied whenever the condition $Na/a_\perp\gg1$ is verified \cite{Dalfovo1999}. 

For dipolar gases there also exists a TF approximation \cite{Dell2004}, and it is possible to obtain semi-analytical expressions, which come in terms of a complicated transcendental relation. In this case the condition $Na/a_\perp\gg1$ does not define the TF regime, since it does not take the dipolar interaction into account. However, a qualitative characterization of the TF limit in dipolar condensates relies on the parabolic density profile of the ground state \cite{Eberlein2005}.

For the sake of clarity, we will specify when we are considering TF results only applicable to $s$-wave condensates (SWTF), or when we are using a more general approach which only relies on the parabolic profile of the condensate and whose results can be applied both to $s$-wave and dipolar cases (TF).

The energy barrier $\Delta E(d,\Omega)$ that has to be overcome to nucleate a vortex is given by the vortex energy in the rotating frame as a function of the vortex distance $d=\sqrt{x_v^2+y_v^2}$ from the symmetry axis. It can be calculated from the energy in the laboratory frame $E(d)$ using
\begin{equation} \Delta E(d,\Omega) = E(d) - \Omega L_z(d)\ ,\label{EqBar} \end{equation}
where $L_z$ is the expectation value of the angular momentum operator $\hat{L}_z$.
To simplify the notation, we consider $\Delta E(d,\Omega)$ and $E(d)$ to be the energies referred to the ground state of the system. Note that the energy barrier depends on the actual value of $\Omega$ but the  energy in the laboratory frame does not. 
For $\Omega=\Omega_c$ and $d=0$ we have $\Delta E(0,\Omega_c)=0$ since under this conditions the energy of a centered vortex is exactly the rotation energy, which comes directly from the definition of critical frequency \cite{Dalfovo1999}. For a vortex near the boundary of the condensate $L_z\rightarrow 0$ \cite{Muntsa2001}, whence $\Delta E(d,\Omega)\rightarrow 0$. Between these two limits, the energy barrier reaches a maximum value $\Delta E_{\text{max}}$ at the position $d_{\text{max}}$, corresponding to the extra energy the system needs in order to carry the vortex from the surface to the center of the condensate.

In the TF limit it is possible to obtain analytic expressions for the energy barrier for vortex nucleation. Although we are dealing with 3D systems, since the trap anisotropy parameter is large, $\lambda=\omega_z/\omega_\perp=11$, it is worth considering the 2D limit. We will derive an analytical expression for 2D systems and briefly discuss the results for 3D condensates \cite{Svidzinsky2000,Kramer2002,Fetter2009}.

\subsection{Two-dimensional case}

In the TF limit the vortex core can be considered to be much smaller than the radius of the condensate, i.e. $\xi\ll R_\perp$, where $\xi$ is the healing length. 
Then, a good approximation is to consider the energy of the vortex to be proportional to the energy of the centered-vortex configuration \cite{Sheehy2004}
\begin{equation} E(d)\simeq\frac{\rho(d)}{\rho(0)}E(0)\ ,\end{equation}
where $\rho(d)$ and $\rho(0)$ are, respectively, the ground state densities at positions $d$ and $d=0$.
Then the energy barrier takes the form
\begin{equation} \Delta E(d,\Omega) = \frac{\rho(d)}{\rho(0)}E(0) - \Omega L_z(d)\ . \label{EqBar} \end{equation}
%
%
%
%
%
To obtain an expression of the barrier, one substitutes $\rho(d)$, $\rho(0)$ and $L_z(d)$ in Eq.~(\ref{EqBar}) by their TF expressions \cite{Muntsa2001}, which assume a parabolic profile of the density of the vortex-free state. This procedure yields
\begin{align}
&\Delta E^{2D}(d,\Omega) = \nonumber\\
& = \Omega_c^{2D}\left[1-\left(\frac{d}{R_{TF}}\right)^2\right]- \Omega\left[1-\left(\frac{d}{R_{TF}}\right)^2\right]^2\ . \label{EqBarTF2D}
\end{align}

Maximizing Eq.~(\ref{EqBarTF2D}), one can find the position of the barrier height
\begin{equation} d_{\text{max}}^{2D} = R_{TF}\sqrt{1-\frac{\Omega_c^{2D}}{2\Omega}}\end{equation}
and its value
\begin{equation} \Delta E_{\text{max}}^{2D}(\Omega)\equiv\Delta E^{2D}(d_{\text{max}},\Omega)=\frac{(\Omega_c^{2D})^2}{4\Omega}\ ,\end{equation}
where the TF radius in the radial direction $R_{RF}$ is related to its root-mean-square ($rms$) analogue $R_\perp$ by $R_{TF}=\sqrt{3}R_\perp$.

In the SWTF limit, one can find an analytical expression for the critical frequency \cite{Lundh1997}
\begin{equation}
 \Omega_c^{2D}= 2\frac{\hbar}{mR_{TF}^2}\ln\frac{0.888R_{TF}}{\xi} \ ,\label{Omega2D}
\end{equation}
which takes the value $ \Omega_c^{2D}=0.34\,\omega_\perp$ for our parameters. To obtain Eq.~(\ref{Omega2D}) an explicit expression for $\xi$ is used which can only be applied to $s$-wave condensates.

\subsection{Three-dimensional case}

In 3D, the procedure to obtain an expression for the energy barrier differs from the 2D case. It has been reported elsewhere \cite{Svidzinsky2000,Kramer2002, Fetter2009}, and here we only summarize the main results. The nucleation energy barrier has the expression
\begin{align}& \Delta E^{3D}(d,\Omega) =\nonumber\\
=& \Omega_c^{3D}\left[1-\left(\frac{d}{R_{TF}}\right)^2\right]^{3/2} - \Omega\left[1-\left(\frac{d}{R_{TF}}\right)^2\right]^{5/2}\label{EqBarTF3D}\end{align}
with the maximum at the position
\begin{equation}d_{\text{max}}^{3D} = R_{TF}\sqrt{1-\frac{3}{5}\frac{\Omega_c^{3D}}{\Omega}}\end{equation}
and a barrier height
\begin{equation}\Delta E_{\text{max}}^{3D}(\Omega)\equiv\Delta E^{3D}(d_{\text{max}},\Omega)=\frac{2}{3}\Omega_c^{3D}\left(\frac{3}{5}\frac{\Omega_c^{3D}}{\Omega}\right)^{3/2}\ .\end{equation}
 As in the 2D case, the relation $R_{TF}=\sqrt{3}R_\perp$ also holds. 

Note that expression (\ref{EqBarTF3D}) has been found only using that the density profile is parabolic, and makes no specific assumption of purely $s$-wave interacting BECs. Therefore, it also applies when dipolar interactions are considered, provided the density profile is parabolic, as in 2D, Eq.~(\ref{EqBarTF2D}).

Again, in the SWTF limit one can find an analytic expression for the critical frequency \cite{Lundh1997}
\begin{equation}
 \Omega_c^{3D}= \frac{5}{2}\frac{\hbar}{mR_{TF}^2}\ln\frac{0.671R_{TF}}{\xi} \ , \label{Omega3D}
\end{equation}
which takes the value $ \Omega_c^{3D}=0.38\,\omega_\perp$ in the present work.




\section{Numerical calculation of the energy barrier}\label{Numerical}

In this section we use the numerical procedure outlined in Sec.~\ref{Theory} to solve the GP equation for different vortex distances $d$ from the symmetry axis. 
An estimate of the formation energy of the vortex is then obtained from the energy in the rotating frame given by the energy functional Eq.~(\ref{ed}) relative to the ground state energy. 

\begin{figure}
\epsfig{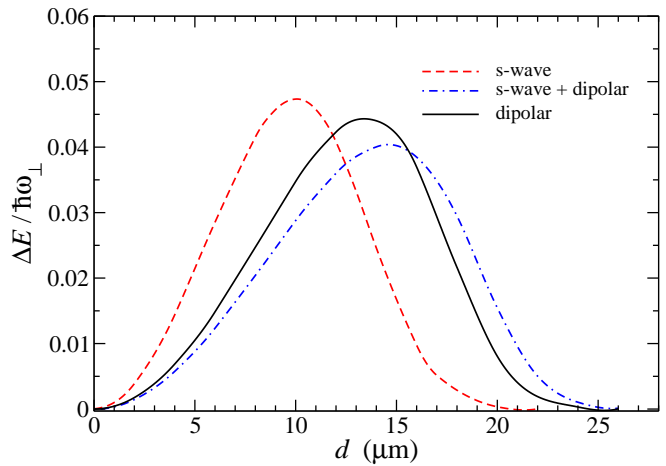}
\caption{(Color online) Energy barrier for the nucleation of a vortex in the
rotating frame at $\Omega=\Omega_c$ as a function of the vortex
displacement from the center. The dashed line corresponds to
the pure contact interaction BEC (with $a=5 a_B$), the solid
line corresponds to a pure dipolar BEC, and the dash-dotted
line to a condensate with contact plus dipolar interactions.
}
\label{barrier-critica}
\end{figure}

We plot in Fig.~\ref{barrier-critica} the vortex formation energy
as a function of the vortex displacement from the center, for the same cases as shown in Fig.~\ref{profiles} and for $\Omega=\Omega_c$. The critical rotation frequency is calculated from thermodynamical arguments \cite{Dalfovo1999, Fetter2009}, giving
\begin{equation}
 \Omega_c=\frac{1}{NL_z}(E(0)- E_0)\ , \label{Omegac}
\end{equation}
where $E(0)$ and $E_0$ are the energies of the centered vortex state and the ground state in the laboratory frame, respectively.
 We see in the figure that the centered-vortex state and the
vortex-free state have the same energy in the rotating frame but they are indeed
separated by an energy barrier. The critical frequencies are different in each case, namely: $\Omega_c=0.45\hbar\omega_\perp$ for a purely $s$-wave condensate, and $\Omega_c=0.25\hbar\omega_\perp$ for a condensate with both interactions and for a purely dipolar condensate.
%

The dashed line in Fig.~\ref{barrier-critica} corresponds to the purely contact interaction
BEC, the dash-dotted line to a condensate with
contact plus dipolar interactions, and the solid
line corresponds to a purely dipolar BEC. 
%
The effect of the strength of the overall interaction is to decrease the barrier height: it is higher for an $s$-wave condensate because it is the weakest interacting case (see Fig.~\ref{profiles}) and smaller for a condensate with dipolar plus $s$-wave interactions, since here the effective repulsion is the strongest. Besides, when dipolar interactions are considered the radial extent of the condensate is larger and thus the barrier height is displaced towards the surface, as compared to the purely $s$-wave case.

\begin{figure}
\epsfig{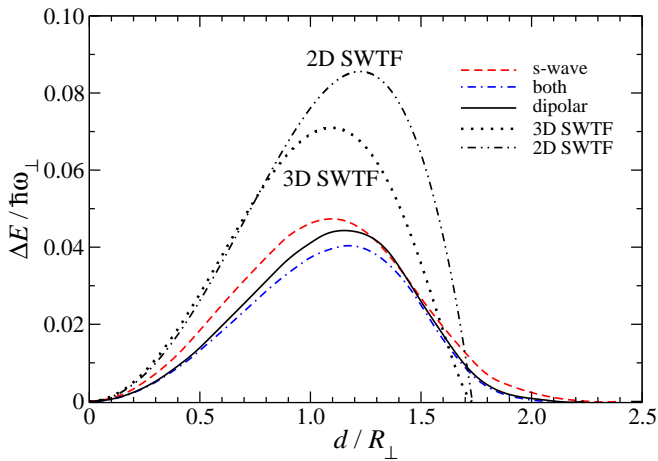}
\caption{(Color online) Energy barrier for the nucleation of a vortex in the
rotating frame at $\Omega=\Omega_c$ for the same cases as in Fig.~\ref{barrier-critica}, together with the predictions of the SWTF model in 2D and 3D.
}
\label{barrier-critica_norm}
\end{figure}

To get rid of the effect of the different sizes due to the different strengths of the repulsive interactions, we plot the energy barriers in Fig.~\ref{barrier-critica_norm} as a function of the vortex displacement normalized to the transversal $rms$-radius of the corresponding ground state, $d/R_\perp$.
It is interesting to note that even though $R_{\perp}$ is different for each curve, the three critical barriers have the same
qualitative behavior when expressed in units of $R_\perp$. 
%

In Fig.~\ref{barrier-critica_norm} we compare the numerical barriers to those obtained from the SWTF results, which correspond to Eqs.~(\ref{EqBarTF2D}) and (\ref{EqBarTF3D}) using the critical frequencies Eqs.~(\ref{Omega2D}) and (\ref{Omega3D}), respectively. 
The disagreement between the SWTF prediction and the numerical calculations can be understood recalling that the parameters used in the present work lead to $Na/a_\perp=5.5$, which is not much larger than 1 and therefore we are not in the full range of validity of the SWTF approximation.

Nonetheless, to obtain an order of magnitude for the energy barriers, it is possible to combine the TF results (\ref{EqBarTF2D}) and (\ref{EqBarTF3D}) with the numerical values of $\Omega_c$ and $R_\perp$ obtained from the GP equation. 
This pseudo-analytical method avoids thus the calculation of the complete energy barrier.
Table~\ref{table} shows the main parameters of the barriers using these approximations at the critical frequency.

\begin{table}[h]\centering\caption{Comparison of the energy barrier maximum and its position obtained in the numerical simulation (sim.) and the 2D and 3D pseudo-analytical results.}\label{table}
 \begin{tabular}{lcc|ccc}
  Case &  $\Omega_c/\hbar\omega_\perp$ & $R_\perp (\mu$m) &  & $\Delta E_{\text{max}}/\hbar\omega_\perp$ & $d_{\text{max}}/R_\perp$ \\\hline\hline 
           &        &        & sim. & $0.047$ & $1.09$ \\
  $s$-wave & $0.45$ & $9.16$ & 2D   & $0.112$ & $1.22$ \\
           &        &        & 3D   & $0.084$ & $1.09$ \\ \hline\hline 
                     &        &         & sim. & $0.040$ & $1.17$ \\
  both & $0.25$ & $12.44$ & 2D   & $0.062$ & $1.22$ \\ 
                     &        &         & 3D   & $0.046$ & $1.09$ \\\hline\hline
           &        &         & sim. & $0.044$ & $1.16$ \\
  dipolar  & $0.25$ & $11.62$ & 2D   & $0.062$ & $1.22$ \\
           &        &         & 3D   & $0.046$ & $1.09$ \\
 \end{tabular}
\end{table}

We see from Table~\ref{table} that the 3D approximation is in good agreement with the characteristics of the barriers found in the numerical simulation, whereas the 2D approximation shows some deviation. This means that in spite of the large anisotropy of the trapping potential, the 3D nature of the condensate is important to characterize the vortex energetics. In addition, the energy of the maximum is better reproduced for condensates with dipolar or both interactions. This happens because the corresponding density profiles (see Fig.~\ref{profiles}) are closer to the TF inverted parabola profile than those of the purely $s$-wave condensate.

%
%
%
%
%

\section{Conclusions}\label{Conclusions}

In this work we have studied the energy barriers for vortex nucleation in the case of dipolar condensates. We have considered the regime of small scattering lengths, which in our case means that dipolar effects are enhanced, and have obtained the vortex formation energies as a function of the vortex displacement from the symmetry axis. We have compared the results obtained when $s$-wave plus dipolar interactions are considered with those obtained in the cases of condensates interacting only via $s$-wave interactions and condensates with only dipolar interactions.

Completely analytic expressions can be found in the $s$-wave Thomas-Fermi regime for the energy barriers in 2D and 3D. From these expressions the position of the barrier maximum and its value can be obtained. Since they mainly hold for $Na/a_\perp\gg1$, their predictions are not in much agreement with our numerical results.

However, a pseudo-analytical approach can be used which is based on the equations from TF theory, but taking the critical frequency and radius of the condensate from the numerical results. This procedure only assumes parabolic density profiles for the ground state densities, thus being applicable to both $s$-wave and dipolar condensates, provided they satisfy the above condition. 

For condensates with dipolar interaction, this pseudo-analytical method has been shown to provide close quantitative results for the nucleation energy barrier when the 3D Thomas-Fermi expression is used. 
Since the density profiles for the dipolar and $s$-wave plus dipolar condensates are approximately parabolic, the Thomas-Fermi expression for the energy barrier holds and its predictions are in accordance with the numerical results. This method can be a useful tool to estimate the energy barrier for vortex nucleation, avoiding thus the calculation of the complete energy barrier.


\acknowledgments We thank Manuel Barranco for helpful discussions. This work has been performed under Grants No. FIS2008-00421 from MEC (Spain), No. 2009SGR1289 from Generalitat de Catalunya (Spain), and No. PICT 31980/05 from ANPCYT (Argentina). 
M. A. is supported by the Comission for Universities and Research of the Department of Innovation, Universities and Enterprises of the Catalan Government and the European Social Fund.

\end{document}